\documentclass[preprint2]{aastex}
\usepackage{color}

\slugcomment{Sunrise early paper}

\shorttitle{Loop statistics}
\shortauthors{Wiegelmann et al.}

\begin{document}


\title{Magnetic loops in the quiet Sun}


\author{\textsc{T.~Wiegelmann$^{1}$,
          S.~K.~Solanki$^{1,8}$,
          J.~M.~Borrero$^{2}$,
          V.~Mart\'\i nez Pillet$^{3}$,
          J.~C.~del Toro Iniesta$^{4}$,
          V.~Domingo$^{5}$,
          J.~A.~ Bonet$^{3}$,
          P.~Barthol$^{1}$,
          A.~Gandorfer$^{1}$,
          M.~Kn\"olker$^{6}$,
          W.~Schmidt$^{2}$,
           \& A.~M.~Title$^{7}$
}}
\affil{
$^{1}$Max-Planck-Institut f\"ur Sonnensystemforschung, Max-Planck-Str. 2, 37191 Katlenburg-Lindau, Germany.\\
$^{2}$Kiepenheuer-Institut f\"ur Sonnenphysik, Sch\"oneckstr. 6, 79104 Freiburg, Germany.\\
$^{3}$Instituto de Astrof\'{\i}sica de Canarias, C/Via L\'actea s/n, 38200 La Laguna, Tenerife, Spain.\\
$^{4}$Instituto de Astrof\'{\i}sica de Andaluc\'{\i}a (CSIC), Apartado de Correos 3004, 18080 Granada, Spain.\\
$^{5}$Grupo de Astronom\'{\i}a y Ciencias del Espacio, Universidad de Valencia, 46980 Paterna, Valencia, Spain.\\
$^{6}$High Altitude Observatory, National Center for Atmospheric Research, Boulder, CO 80307, USA.
The National Center for Atmospheric Research is sponsored by the National
Science Foundation.\\
$^{7}$Lockheed Martin Solar and Astrophysics Laboratory, Bldg. 252, 3251 Hanover Street, Palo Alto, CA 94304, USA.\\
$^{8}$School of Space Research, Kyung Hee University, Yongin, Gyeonggi, 446-701,
Korea.\\
}
\email{wiegelmann@mps.mpg.de}




\begin{abstract}
We investigate the fine structure of magnetic fields in the
 atmosphere of the quiet Sun.
We use photospheric magnetic field measurements from {\sc Sunrise}/IMaX with
unprecedented spatial resolution to extrapolate  the photospheric magnetic field
into higher layers of the solar atmosphere
with the help of potential and force-free extrapolation techniques.
We find that most magnetic loops which
 reach into the chromosphere or higher have  one foot point
 in relatively strong magnetic field regions in the photosphere.  $91 \%$ of the
 magnetic energy in the mid chromosphere (at a height of 1 Mm) is
 in field lines, whose stronger foot point has a strength
of more than 300 G, i.e. above the equipartition field strength with
convection. The loops reaching into the chromosphere and corona
are also found to be asymmetric in the sense that the weaker
foot point has a strength $B < 300$ G and is located in the internetwork.
Such loops are expected to be strongly dynamic and have short lifetimes,
as dictated by the properties of the internetwork fields.
\end{abstract}
\keywords{Sun: magnetic topology---Sun: chromosphere---Sun: corona---Sun: photosphere}
\section{Introduction}
The Sun's magnetic field lies at the heart of the heating of the solar
corona and the solar chromosphere \citep[with the possible exception of the
basal flux; see, e.g.,][]{bello:etal10}. Unfortunately, however, the magnetic field is measured almost
exclusively in the solar photosphere and needs to be extrapolated from there in
order to obtain  its structure in the Sun's upper atmosphere.
The balloon-borne {\sc Sunrise} mission
\citep{solanki:etal10, barthol:etal10}
obtained the
magnetic field in the quiet solar photosphere with a very high
and homogeneous spatial
resolution. This allows us to investigate
the 3D structure of the quiet Sun's
magnetic field in more detail as compared with SOHO
\citep[e.g.][]{wiegelmann:etal04,he:etal07} or Hinode
\citep[e.g.][]{martinez:etal09,martinez:etal10}.

 The resolution of the extrapolated field
in the vertical direction depends on the spatial scales of the
photospheric measurements. With a pixel size of 40 km on the Sun
and a spatial resolution of $\sim 100$ km,
we can, for the first time, resolve the thin layer of the photosphere and
the lower chromosphere with several grid points. This allows us
to study the magnetic connectivity between photosphere, chromosphere
and corona by well-resolved magnetic loops.
We also briefly discuss the possible implications of our
results for quiet Sun loop heating models. This is of particular
interest due to the large number of horizontal magnetic features
in the photosphere, many of them associated with emerging magnetic
loops \citep{martinez:etal08, danilovic:etal10} and
the lack of acoustic wave energy flux proposed by
\cite{carlsson:etal07} which inspires us to consider alternatives, although the
recent work by \citet{bello:etal09,bello:etal10} suggests that
the acoustic energy flux may well have been underestimated in the
past due to insufficient spatial resolution.
\section{Extrapolation of photospheric magnetic field measurements into the upper solar atmosphere}
\begin{figure*}
$ $ \\
\includegraphics[width=8cm]{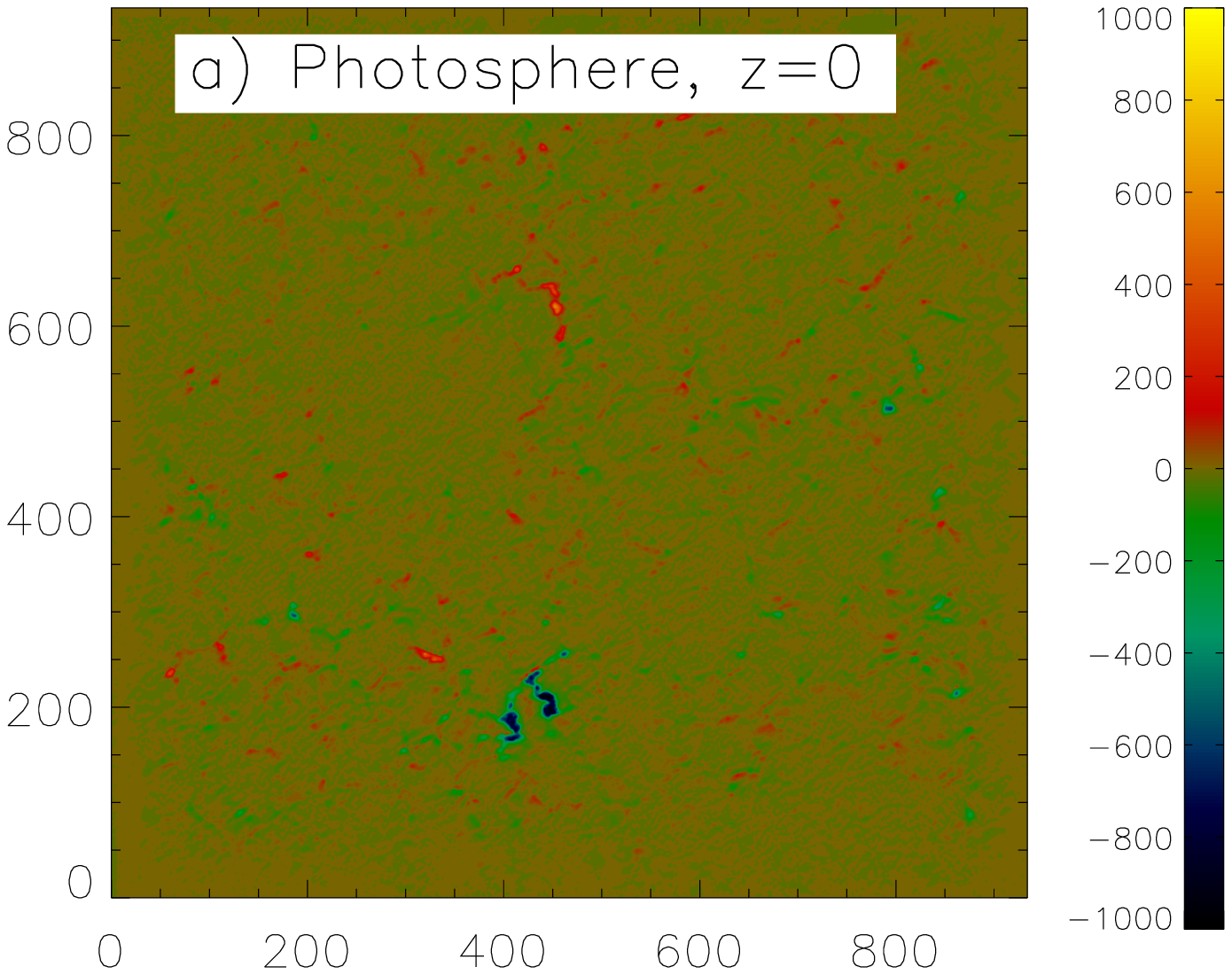}
\includegraphics[width=8cm]{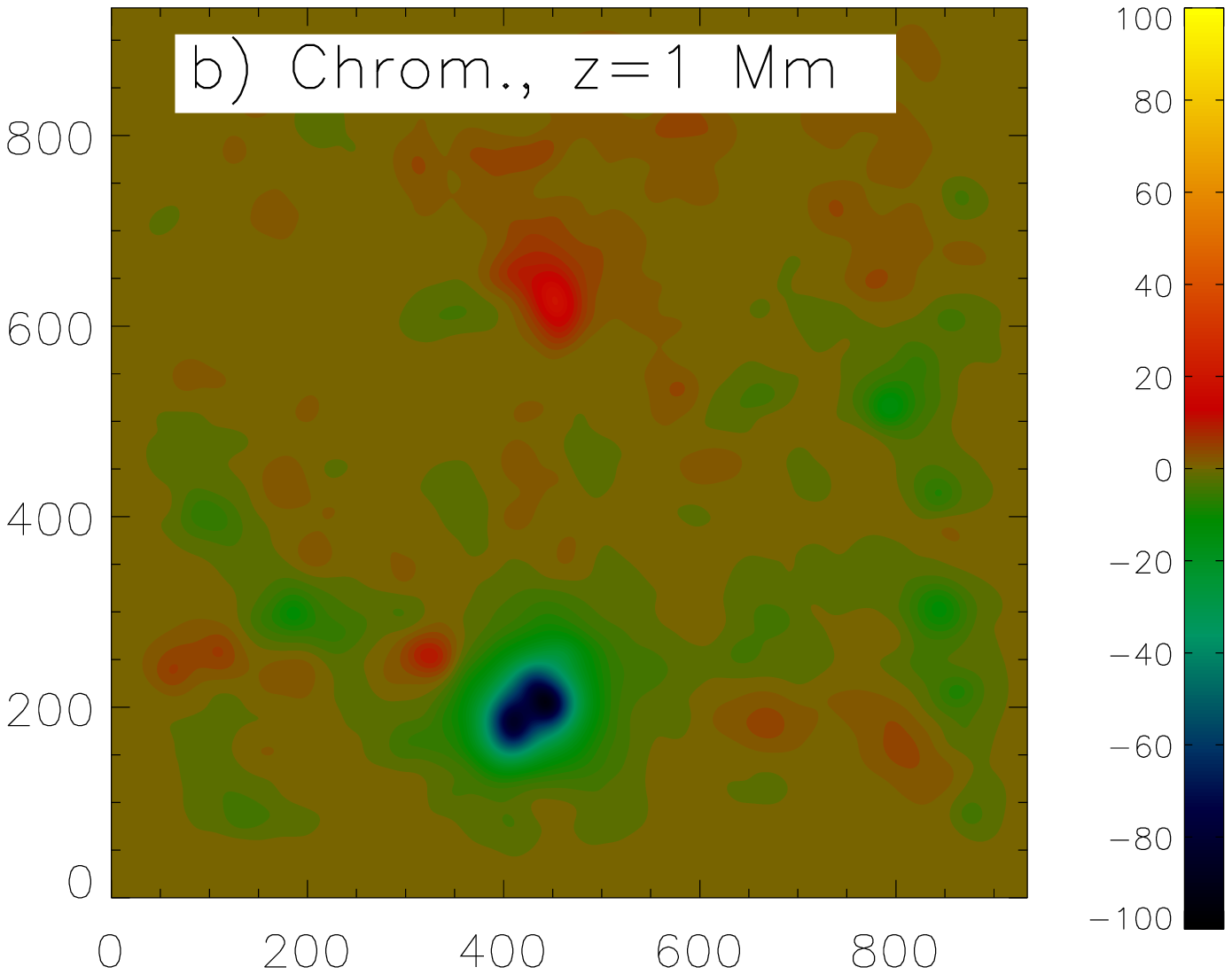}
\includegraphics[width=8cm]{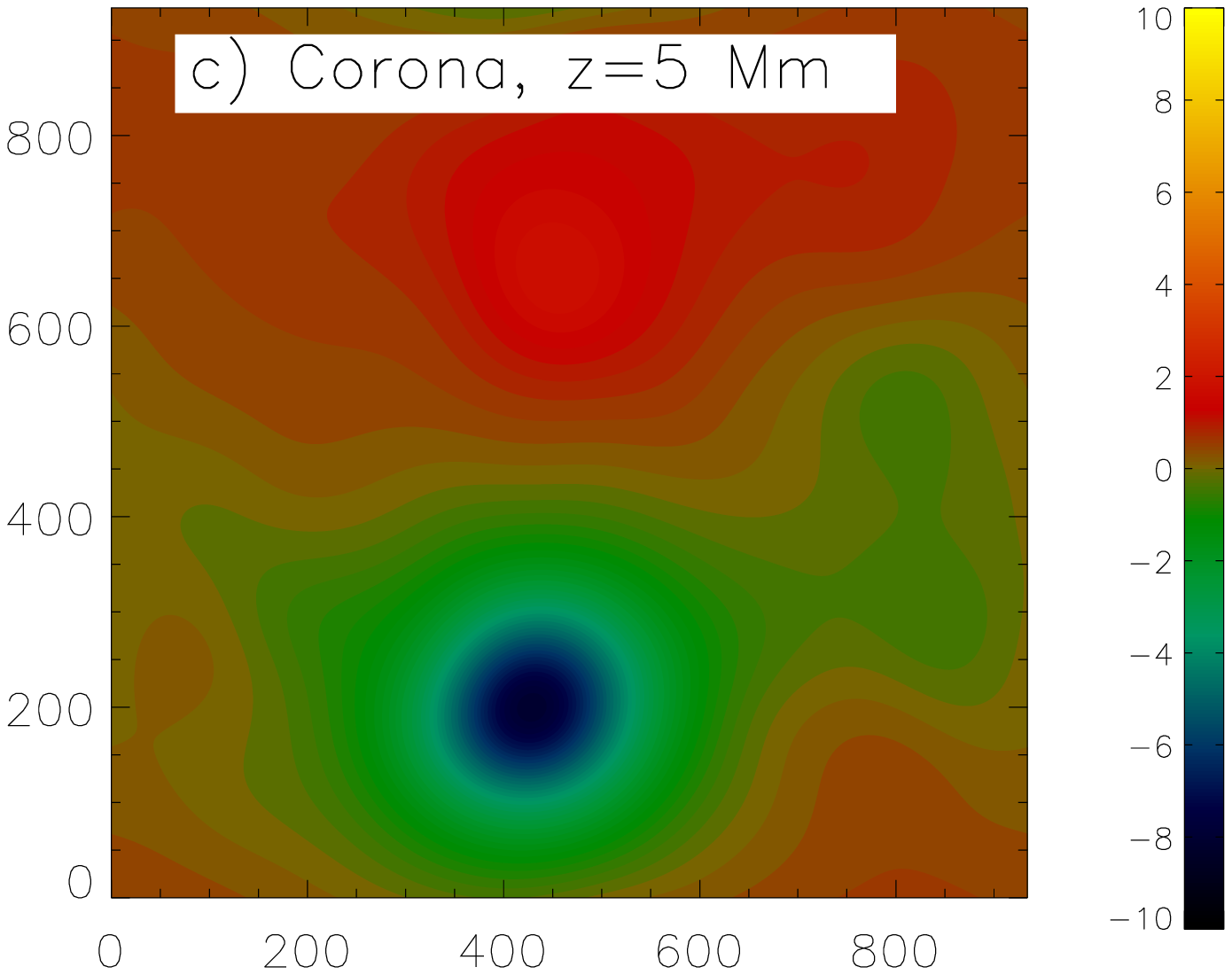}
\includegraphics[width=8cm]{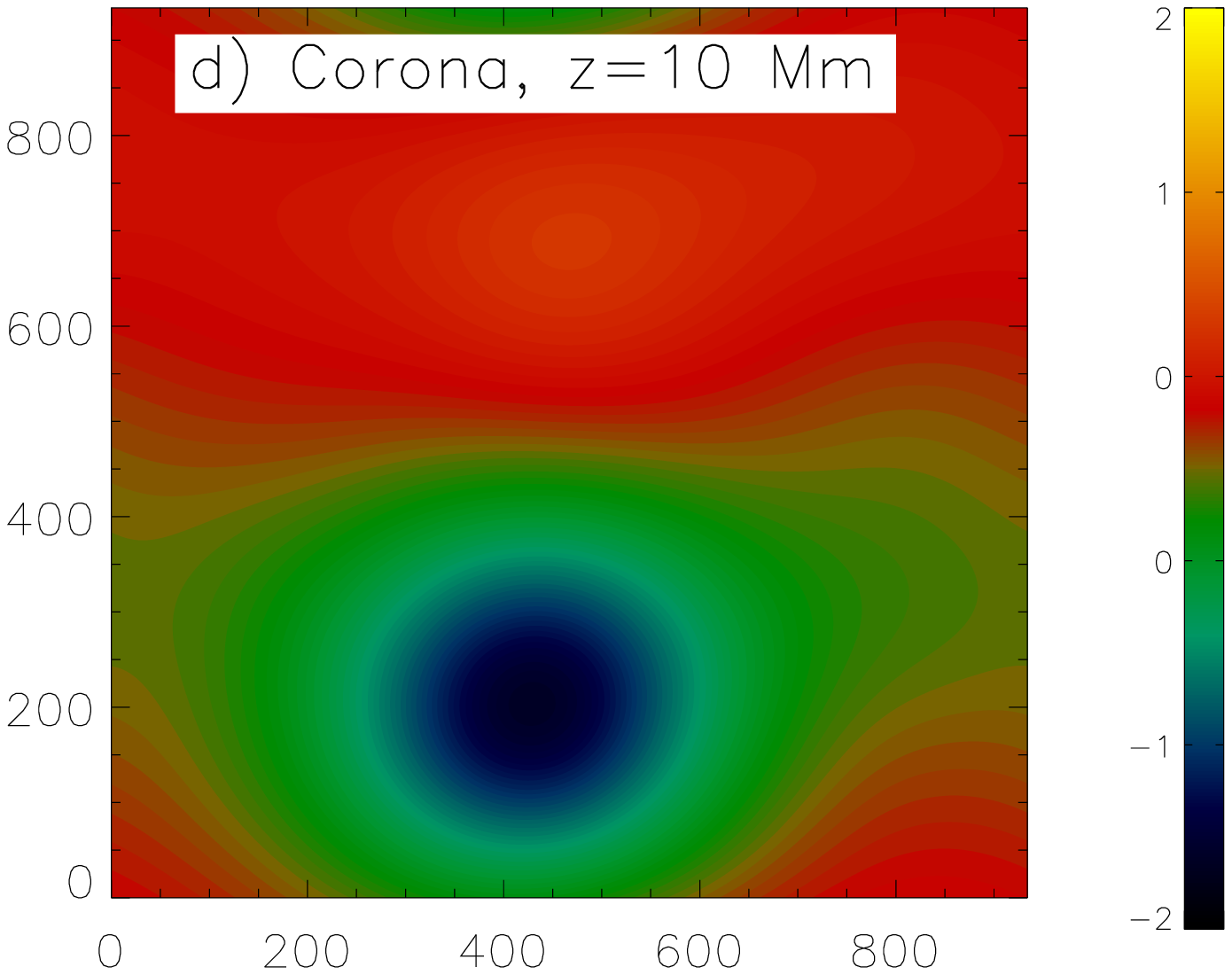}
$ $ \\
\caption{Vertical magnetic field  at different heights. Panel a) shows the measured
magnetogram in the photosphere and panels b-d) the vertical magnetic field distribution in different
heights in the solar atmosphere, as reconstructed with a potential field model. The colour bars
show the magnetic field strength in G.}
\label{fig1}
\end{figure*}
Here we use a (phase diversity reconstructed)
 Stokes vector map of a $37 \times 37$ Mm quiet Sun region in
 the photosphere recorded by {\sc Sunrise}/IMaX
\citep[][the data set was observed for $1.616$ hours starting
at 00:00 UT on 2009 June 9th.]{martinezpillet:etal10}.
The magnetic vector is obtained
by inverting these data using the VFISV code by \citet{borrero:etal10}.
We take 200 km as the average height at which the Fe {\sc i} spectral
line at $5250.2$ \AA\ senses
$B$, although it is difficult to define \citep[][]{sanchez:etal96}
and, in fact, varies from point to point over the solar surface.
The height of this layer corresponds to $z=0$ in Fig. \ref{fig1}.

To get a first impression of the 3D magnetic field structure in
the chromosphere and corona we extrapolate the photospheric measurements
into the atmosphere under the force-free assumption.
\begin{eqnarray}
\nabla \times {\bf B } & = & \alpha {\bf B}  \label{eq1}\\
 \nabla\cdot{\bf B}    & = &   0      \label{eq2}\\
 {\bf B} & = & {\bf B}_{\rm obs}\; {\rm in \; photosphere}, \label{eq3} 
\end{eqnarray}
where ${\bf B}$ is the magnetic flux density and $\alpha$ is the
force-free parameter. Here we use mainly potential field
extrapolations $(\alpha=0)$ due to the generally
low signal level in the quiet Sun, in particular of the linear polarization.
Taking into account the non-force-free character of
photosphere and lower chromosphere
\citep[as, e.g., proposed in][]{petrie:etal00,wiegelmann:etal06b}
require  information on the plasma density and temperature distribution.
The employed approximation is  supported by
spectro-polarimetric observations from
\cite{martinez:etal10}, which suggest that quiet sun loops show a potential
field like structure.
We  solve equations (\ref{eq1})-(\ref{eq3})
with the help of a fast Fourier approach
\citep[see][]{alissandrakis81} with the measured vertical magnetic field as
boundary condition.
A Fourier approach is justified because the  magnetogram is
almost flux balanced with $ \sum\limits_{x,y}^{} B_z / \sum\limits_{x,y}^{} |B_z|=-0.077$.
The computations are carried out in
a cubic box with $936 \times 936 \times 468$ grid points in $x,y,
z$, where $z$ is the vertical direction. We use a constant grid
resolution of 40 km in each direction, which corresponds to a total
3D model volume of $37 \times 37 \times 19$ Mm.
Figure \ref{fig1}a shows the line-of-sight photospheric magnetic field and
slices of the reconstructed magnetic field at different heights
in the solar atmosphere
in Fig. \ref{fig1}b-d. With increasing height in the solar atmosphere the magnetic field
becomes smoother and shows a dipolar structure at coronal heights.
\section{Statistics of magnetic loops.}
\begin{figure}
\includegraphics[width=8cm]{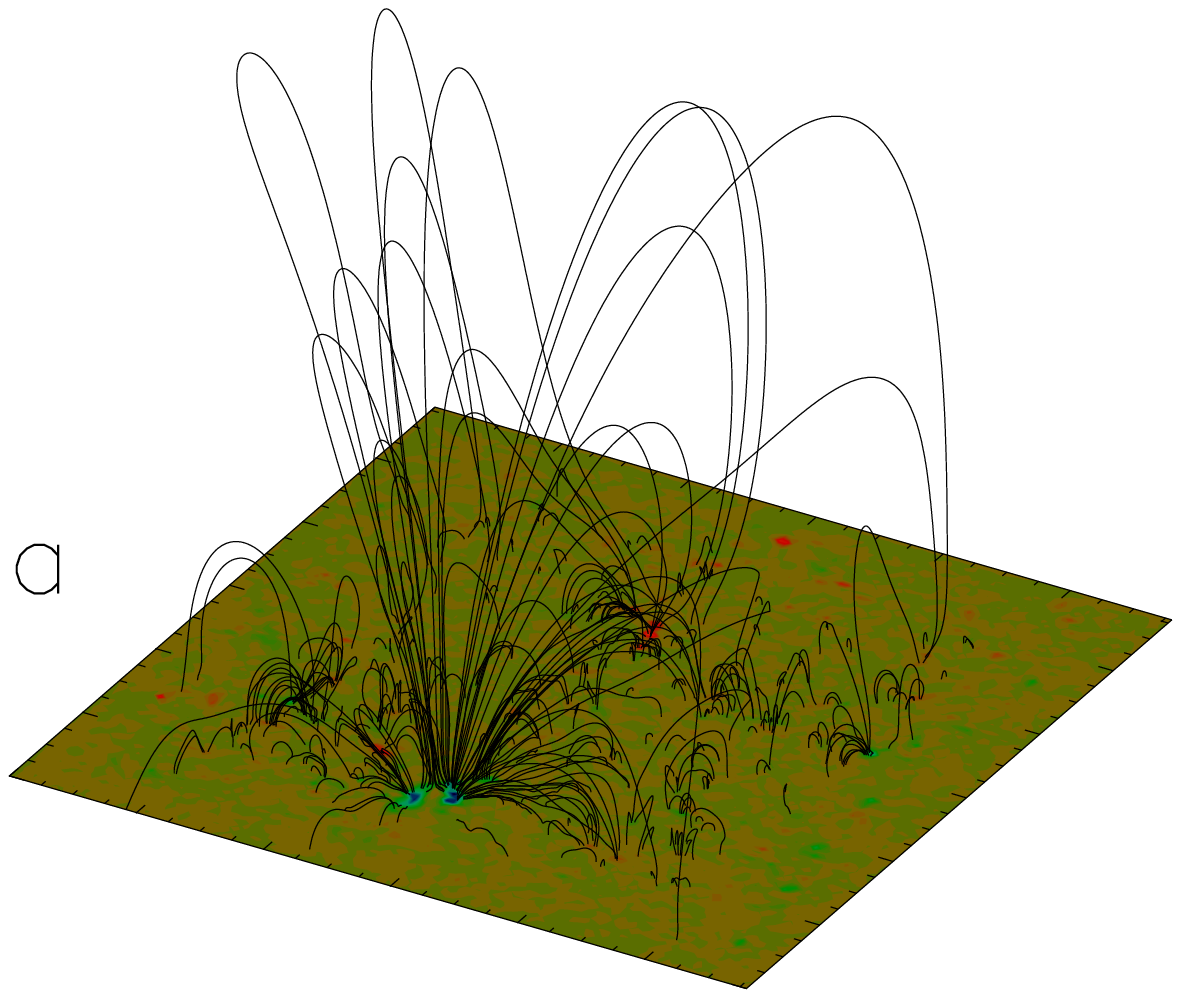}
\includegraphics[width=8cm]{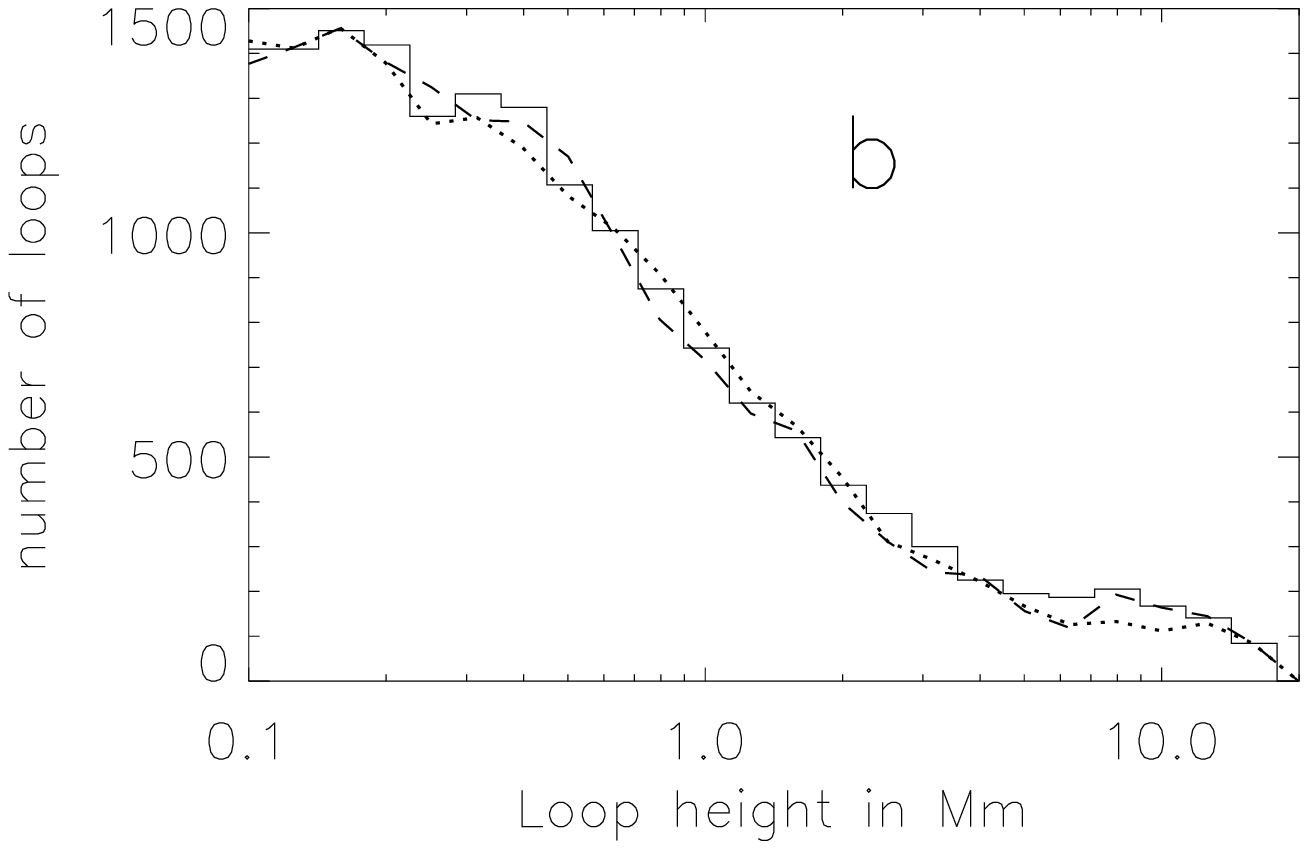}
\includegraphics[width=8cm]{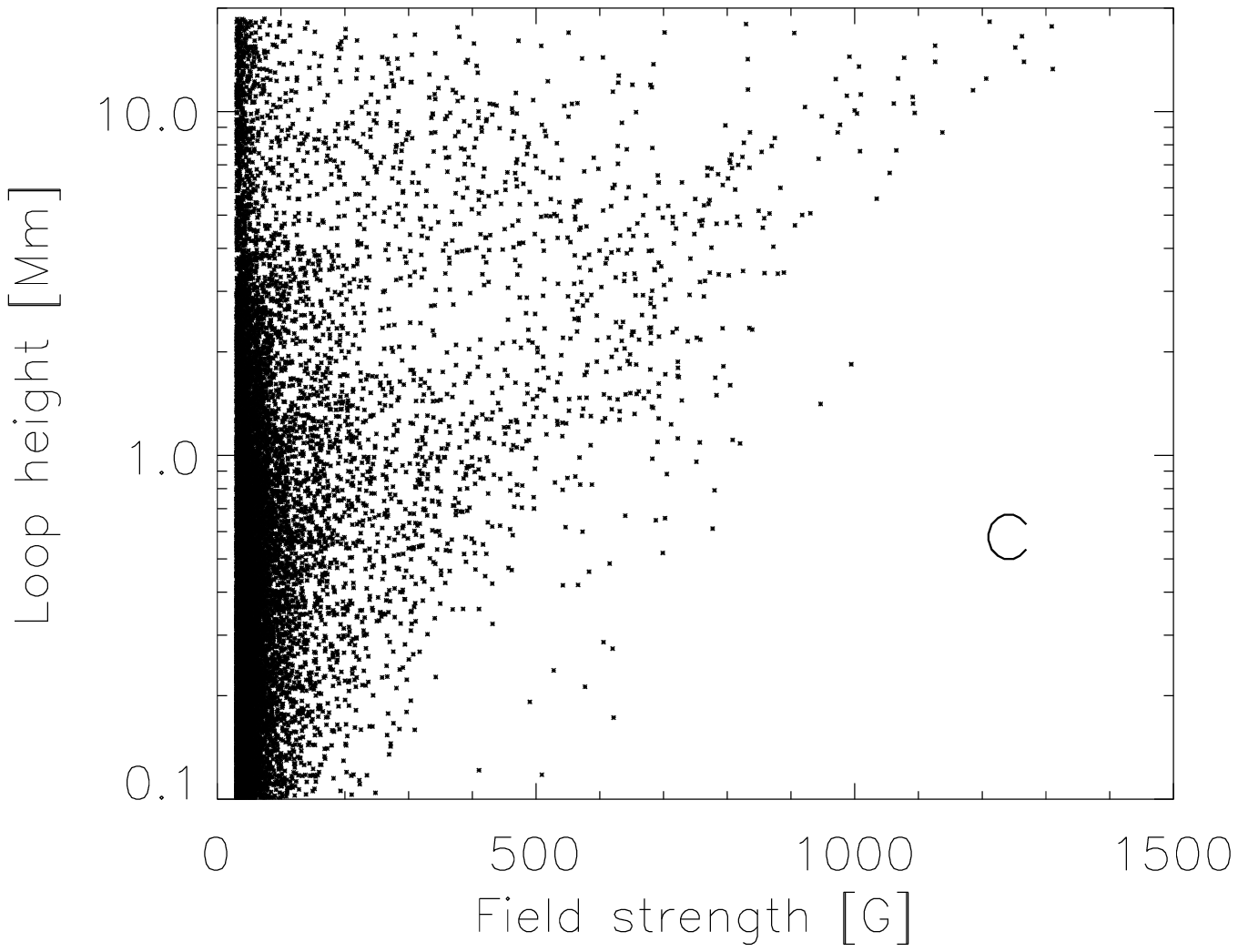}
\caption{a) A random selection of $2 \%$ of magnetic loops, computed from a potential
field reconstruction.
b) Loop statistics for all  resolved loops.
The solid histogram-style line corresponds to a potential field and the
dotted and dashed lines to a linear force-free model with
$\alpha \cdot L= 4 \, {\rm and} -4$, respectively.
c) Scatter plot of loop height vs. leading foot point vertical field.}
\label{fig2}
\end{figure}
We compute magnetic field
lines from all pixels above a certain threshold value, here $|B_z| >
30$ G in the photosphere, which corresponds to the $3 \sigma$ noise level
of the magnetogram.
The field line integration is started in a
rectangular area at the center of the magnetogram, $150$ pixels or
$6$ Mm away from the lateral boundaries. A total of
28005 field lines are followed from one foot point to the other. Of
these $1936 \, (7 \%)$ leave the
computational domain through the lateral or top boundaries. We
cannot say if these lines are really open or close outside the
{\sc Sunrise}/IMaX FOV. Consequently, we do not consider them further
in this section.
For simplicity we call closed magnetic field lines {\it loops} for
the rest of this paper. Such magnetic loops are ubiquitous in the solar
atmosphere, although only a fraction of these loops might be visible
in images made in chromospheric, transition region or coronal radiation.
Loops with a height of less than $100$ km ($9321, \, 33 \%$ of all field lines),
may not be spatially sufficiently resolved
 and are not considered further,
leaving 16748 field lines ($60 \%$ of all) to be studied.

Figure \ref{fig2}a shows a randomly chosen fraction of $2\%$
of these loops and Fig.  \ref{fig2}b a histogram
 of the loop height distribution for all resolved loops. The average loop
height is $1.24 \pm 2.45$ Mm
\footnote{For higher (lower) threshold values of $|B_z|$ the number of small
loops decreases (increases), which affects the average loop height, e.g.,
to $1.38$Mm and $1.09$Mm for a threshold of 40G and 20G, respectively.}.
The number of small loops (below about 1 Mm) is significantly larger
than that of higher loops.
$51 \%$ of the resolved loops~\footnote{All loops higher than $100$ km,
including the ones not closing within the Sunrise/IMaX FOV.}
close within phototospheric heights
$(<500 \, {\rm km})$ and $29 \%$ in the chromosphere $(500-2500 \, {\rm km})$.
Finally $20 \%$ of the loops reach into the corona $(>2500 \, {\rm km})$,
but half of these long coronal loops do not close within the
Sunrise FOV.

We also investigated how using a linear force-free model instead of
a potential field model influences the loop height statistics.
These models contain the force-free parameter $\alpha$, which cannot be deduced
from the available data. Consequently one unique  linear force-free model cannot be
computed. To investigate the influence of linear electric currents, we
carried out computations with $\alpha L= \pm 4$, respectively, where $L=37$ Mm
is the size of the employed portion of the magnetogram
(see dotted and dashed lines in Fig. \ref{fig2}b, respectively).
  $4$ is already quite a large value for the normalized
force-free parameter $\alpha L$ and close to the mathematical
allowed maximum value
of $|\alpha L|= \sqrt{2} \pi$ \citep[see discussion in][for a
suitable rank of $\alpha$-values for linear force-free
models]{seehafer78}.
As visible in Fig. \ref{fig2}b the loop height statistics are almost
identical to those obtained from potential fields. Small differences
occur only for field lines higher than about 5Mm. For our study, which
mainly concentrates on lower chromospheric loops, it is therefore justified to
consider only potential fields.

Figure \ref{fig2}c contains a scatter plot of loop height
vs. magnetic field strengths at the leading foot point, i.e. the
foot point with the largest field strength.
It shows that loops of any height can begin from weak fields, but only long
loops start from strong-field foot points. This
means that the stronger field  $(|B_z| {^>_{\sim}} 800 \, {\rm G})$
loops are all long enough to
reach into the solar corona, whereas the apex of a loop originating in a weaker
field region can lie in the photosphere, chromosphere or corona.
\section{On the origin of chromospheric magnetic fields and magnetic energy}
\begin{figure}
\includegraphics[width=10cm]{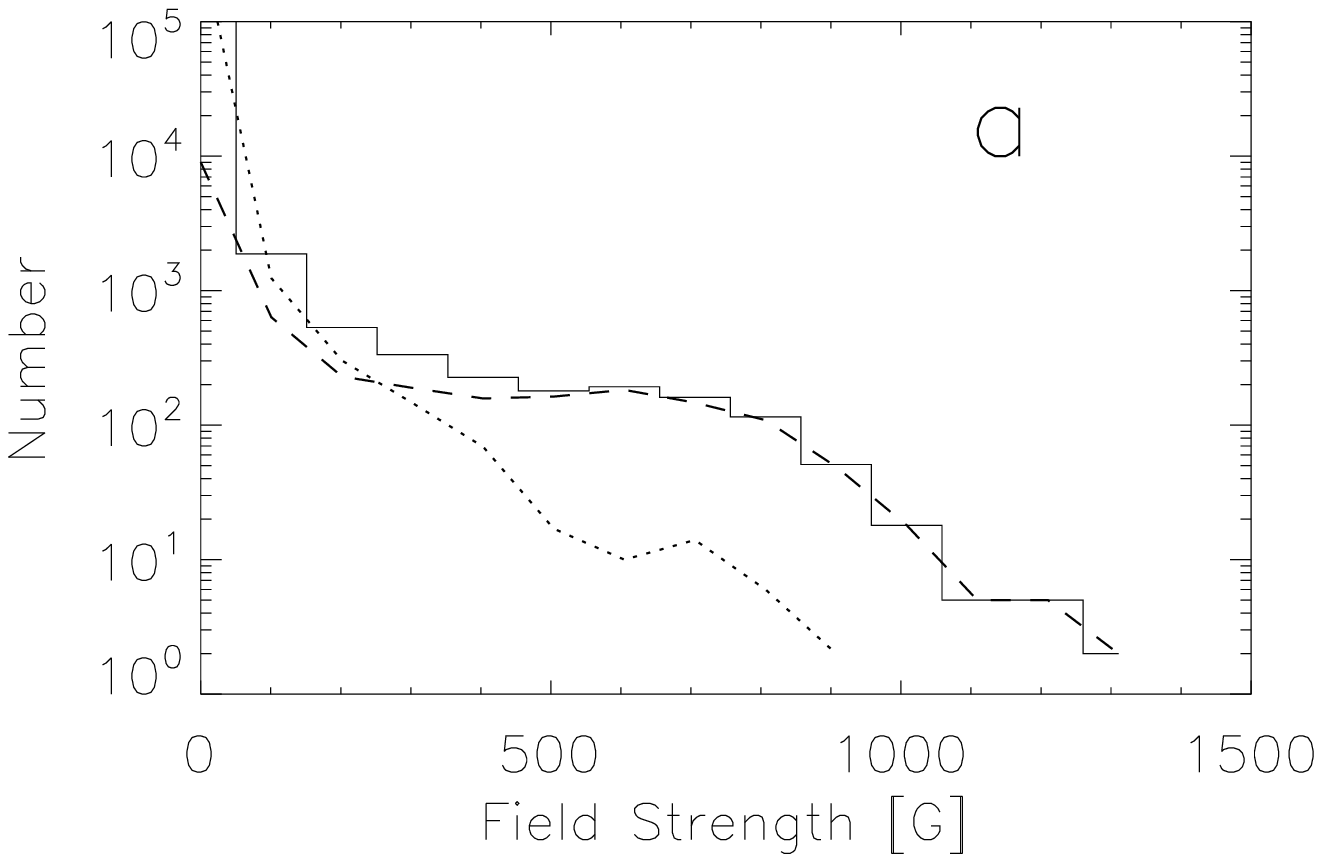}
\includegraphics[width=10cm]{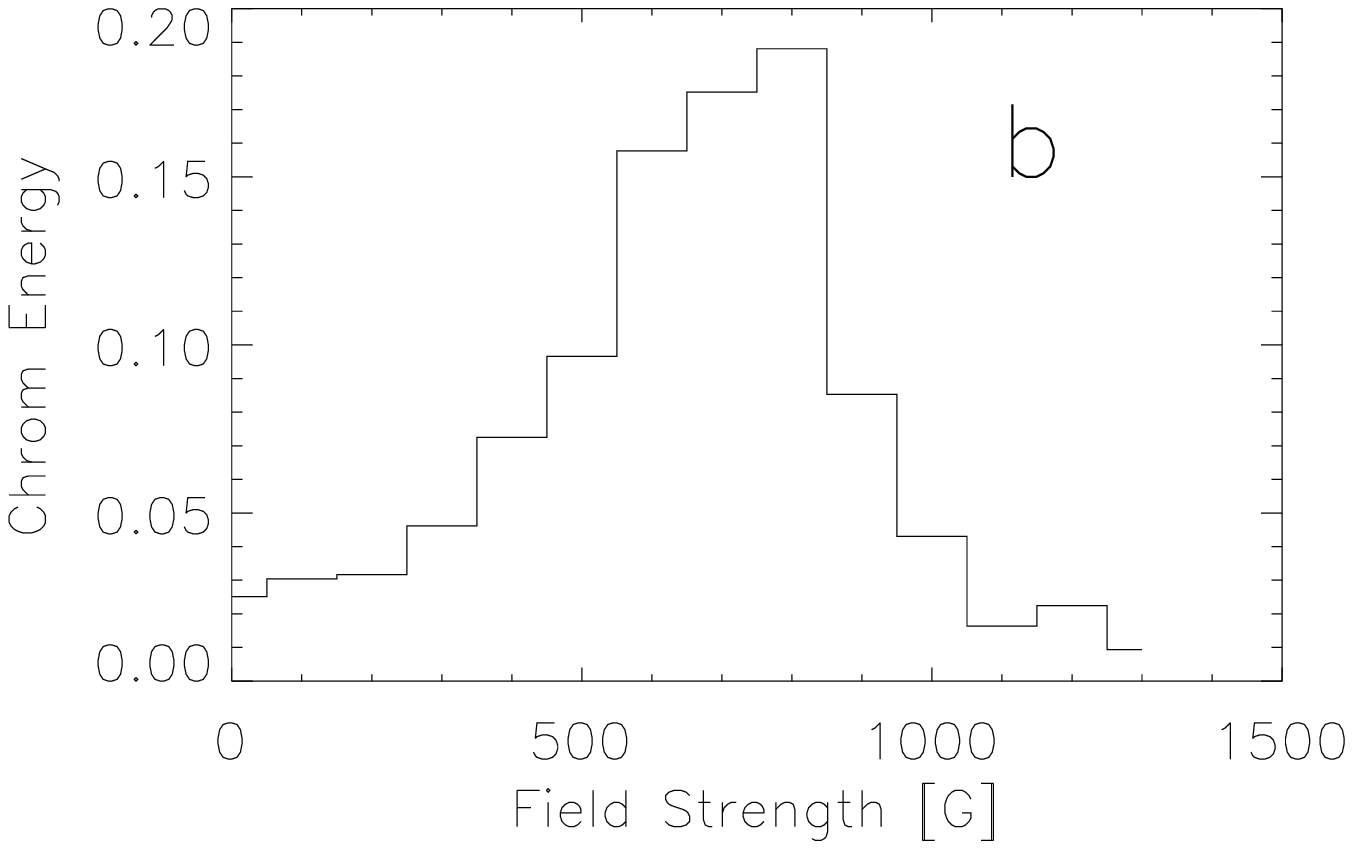}
\caption{a) Photospheric field strength $|B_z|$ statistics of the photospheric
magnetogram (excluding the 150 pixel layer towards the boundaries). The solid
histogram-style line corresponds to all photospheric pixels, the dashed line
to pixels hosting foot points of field lines reaching at least to 1 Mm, i.e. into
the mid-chromosphere, and the dotted line to pixels not hosting such field lines.
b) Histogram showing the fraction of magnetic energy at a height of 1 Mm
(mid-chromosphere) in loops with a particular field strength of their leading
foot point.
}
\label{fig3}
\end{figure}
 In the following we investigate in more detail the origin of magnetic
fields in the mid-chromosphere at a height of 1 Mm
and are in particular interested in the
magnetic connectivity with photospheric fields. To this aim we
start the magnetic field line integration in the chromosphere (see
Fig. \ref{fig1}b) at every pixel except in a layer of $150$
pixels towards the magnetogram boundaries. We track all field lines
down to the photosphere and determine here
in particular the leading, stronger photospheric foot points~\footnote{Some long field lines
do not close within
the Sunrise-FOV and for them we can only identify one foot point.
For the analysis in this section these field lines are included in the statistics
assuming that the known foot point is the stronger one. The relation
to the weaker foot points is investigated in section
\ref{sec_footpoints}.}. From this foot point map we can then
distinguish between photospheric pixels hosting  magnetic loops
reaching into the chromosphere or higher
and pixels not hosting such loops. Figure \ref{fig3}a shows
the magnetic field strength distribution $|B_z|$ in the photosphere.
The solid line is for all pixels, the dashed line for pixels hosting
foot points of field lines that reach to a height of at least $1$ Mm and the
dotted line for pixels not hosting such loops.
Above 500G the solid and dashed lines are practically identical and nearly
all pixels above 500G host footpoints of chromospheric loops.
For weak fields ($< 100 {\rm G}$) only about $2 \%$ of the pixels
host chromospheric loops, but the total number of these (hosting)
pixels is still more than one order of magnitude higher than  strong
field pixels. This raises the question, which photospheric fields can
influence the chromospheric and coronal gas more strongly, the few
regions with  strong fields or the many regions with weak fields?
To answer this question we compute the magnetic energy at
a height of $1$ Mm and investigate which fraction of the energy is
magnetically connected to photospheric foot points with a particular
field strength. As one can see in Fig. \ref{fig3}b the weak
photospheric fields  ($< 100 {\rm G}$) hardly contribute to chromospheric
magnetic energy. In addition, although only $0.91 \%$ of all photospheric
pixels have a field strength above $100$ G, they are connected to $97 \%$ of the
chromospheric magnetic energy. The $0.32 \%$ of pixels above
the equipartition field strength of $300$ G
contribute to $91 \%$ of the chromospheric magnetic energy.
At the equipartition field strength
\citep[which is in the range of 200-400G in the photosphere, see][]{solanki:etal96}
the magnetic energy is identical to the energy of the convective flow. The
equipartition field strength has relevance for quiet Sun magnetic features,
because magnetic flux tubes need to have a larger field strength in
order to survive as a distinct feature.
It is therefor natural to distinguish network elements from
internetwork features by the equipartition field strength. Here we used the
average value of 300G.

Note that the photospheric field strength values are based on single
component inversions assuming spatially resolved fields and are
consequently lower limits
\citep[for the stronger field, they are close to the true
value, see][]{lagg:etal10}, and so are the above energy fractions
of the chromospheric magnetic energy.
The above numbers refer to the leading (stronger) foot point.
\section{Relation between the two foot points of magnetic loops}
\label{sec_footpoints}
\begin{figure}
\includegraphics[width=9cm]{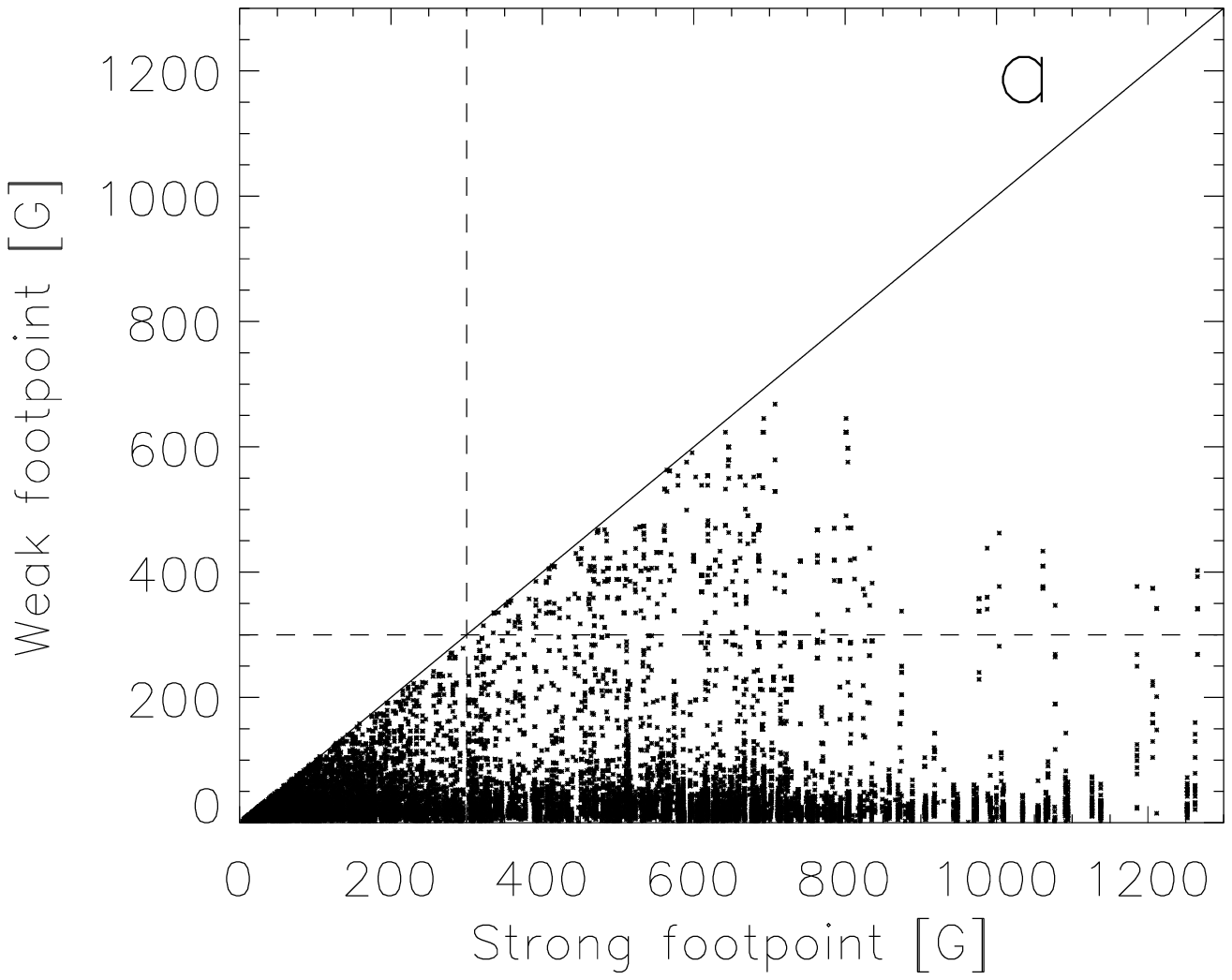}
\includegraphics[width=9cm]{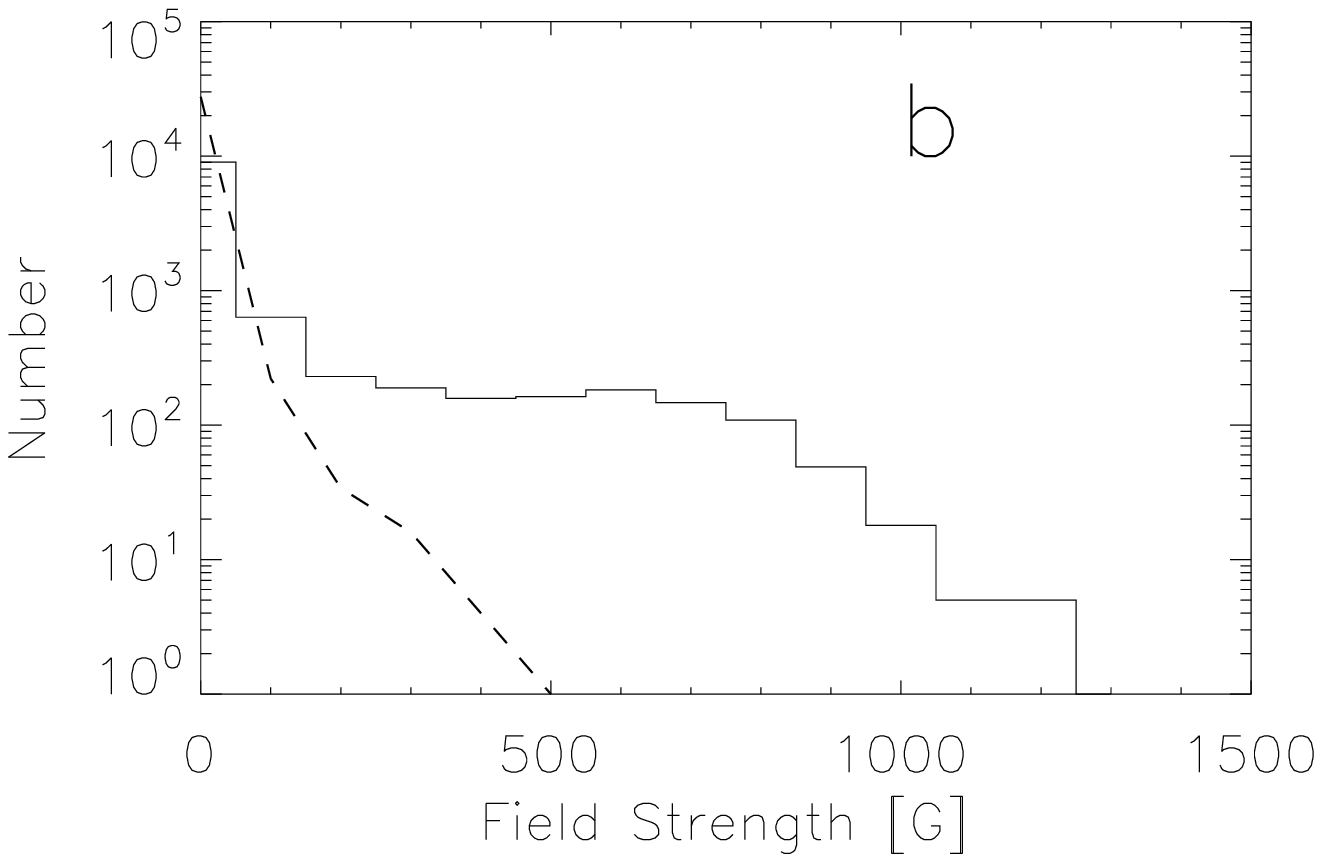}
\caption{Relation between strong and weak foot points of magnetic loops which
reach at least into the chromosphere.
 Excluded have been loops that do not close within the Sunrise-FOV ($14 \%$),
because the strength of one foot point remains unknown.
Panel a) shows a scatter plot of weak foot point strength vs. strong foot
point strength. The solid diagonal line corresponds to equal strength of
both foot points and the dashed horizontal and vertical lines mark the
eqipartition field strength of $300$ G.
Panel b): Photospheric field strength $|B_z|$ statistics for the
strong (solid line) and weak foot points (dashed line) of chromospheric
and coronal loops.
}
\label{fig4}
\end{figure}
In the last section we learned that  $91 \%$ of the mid-chromospheric
magnetic energy is topologically connected to photospheric foot
points with superequipartition
field strength. However, still unknown is if both foot points of
long chromospheric and coronal loops have similar field strength
or if it can be very different. If the latter is the case this
would also mean that the multiple field lines starting within the
area of a strong magnetic pixel will end in many,
possibly widely distributed weak-field
pixels~\footnote{In order to fulfill $\nabla \cdot {\bf B}=0$
we get the relation $B_1 A_1=B_2 A_2$ for the foot point areas $A_1$
and $A_2$ of magnetic flux tubes containing a bundle of field
lines.}.
In Fig. \ref{fig4}a we show a scatter plot
of weak foot point strength vs. strong foot point
strength.~\footnote{Necessarily, we had to exclude those
field lines which do not close within the Sunrise-FOV,
because  we cannot identify both foot points for them.}.
$95 \%$ of all loops with a
leading  foot strength of $>300$ G, have the other foot point in regions
with $<300$ G.
Figure \ref{fig4}b shows the photospheric vertical field
$|B_z|$ statistics for pixels hosting a leading (stronger)
foot point (solid histogram-style line) and for pixel hosting
a weak foot point (dashed line).
Obviously the vertical field
of both foot points for long loops (reaching at least into
the chromosphere) is very different.
Partly, this has to do with the different field strength
distribution in the two magnetic polarities for this particular magnetogram.
The maximum vertical field  in the negative field
region is $1311$ G, compared to  $707$ G
in positive polarity regions \footnote{Indirect evidence of large network
fields of 1-2 kG concentrated in small regions
has been given already in \cite{tarbell:etal77}, but the magnetograph
resolution of that time has been too low to resolve such structures.}.
Investigations of further quiet Sun regions with high resolution are necessary to see if
such different field strengths are common or an exception
\citep[For quiet Sun investigations of magnetic flux see, e.g.][]{wang:etal95}.

\section{Conclusions and outlook}
\label{conclusions}
We investigated the small-scale structure of magnetic fields in the
solar atmosphere with emphasis on the magnetic
connectivity between the photosphere and chromosphere. We found that
chromospheric areas containing about  $91 \%$
of the magnetic energy
are topologically connected to a photospheric foot point with a
strength above $300$ G, i.e. with an energy density higher than
the average equipartition with that of convective flows.
For the majority of the loops the magnetic
field strength in both foot points differs significantly,
with the second foot point having a field strength below
equipartition, i.e. network elements connect magnetically mainly
to internetwork (IN) features, in general agreement with
numerical experiments of \cite{schrijver:etal03a,jendersie:etal06}.
An interesting question
is to which extent the finding that network elements are connected
mainly with IN features
 might influence models of quiet-Sun loop heating
\citep[e.g.,][]{hansteen93,chae:etal02,mueller:etal03,mueller:etal04}.
Such loop heating models usually assume a constant cross section,
an approximation which is not fulfilled for field lines connecting
strong and weak field regions in the photosphere.
It is also worth noting that IN fields are more dynamic and short-lived
than network fields
\citep[IN field average lifetimes are about 10 min according
to][]{dewijn:etal08}.
Since most chromospheric and coronal loops, at least in the observed region, are connected
at one foot point with IN fields we expect the
chromospheric and coronal field
to be more dynamic than suggested by network fields alone. Time series
of high resolution magnetograms in the quiet Sun can be used to
investigate this further and to revisit {\it coronal recycling}
\citep[e.g., as investigated at much lower spatial resolution by][]{close:etal04,close:etal05}.

\begin{acknowledgements}
   The German contribution to $Sunrise$ is funded by the Bundesministerium
   f\"{u}r Wirtschaft und Technologie through Deutsches Zentrum f\"{u}r Luft-
   und Raumfahrt e.V. (DLR), Grant No. 50~OU~0401, and by the Innovationsfond of
   the President of the Max Planck Society (MPG). The Spanish contribution has
   been funded by the Spanish MICINN under projects ESP2006-13030-C06 and
   AYA2009-14105-C06 (including European FEDER funds). The HAO contribution was
   partly funded through NASA grant number NNX08AH38G.
   This work has been partially supported by the WCU grant No. R31-10016
   funded by the Korean Ministry of Education, Science and Technology.
\end{acknowledgements}





\end{document}